\documentclass[%
superscriptaddress,
preprint,
 amsmath,amssymb,
aps,
]{revtex4-2}

\usepackage{filecontents}
\usepackage{graphicx}
\usepackage{dcolumn}
\usepackage{threeparttable}
\usepackage{xr} 
\usepackage{bm}

\begin{document}


\title{Dielectric-Dependent Range-Separated Hybrid Functional Calculations for Metal Oxides}

\author{Jiawei Zhan}
\affiliation{Pritzker School of Molecular Engineering,
University of Chicago, Chicago, Illinois 60637, United States}
\author{Marco Govoni}
\affiliation{Pritzker School of Molecular Engineering,
University of Chicago, Chicago, Illinois 60637, United States}
\affiliation{Materials Science Division and Center for Molecular Engineering, Argonne National Laboratory, Lemont, Illinois 60439, United States}
\affiliation{Department of Physics, Computer Science, and Mathematics, University of Modena and Reggio Emilia, Modena, 41125, Italy}
\author{Giulia Galli}
\email{gagalli@uchicago.edu}
\affiliation{Pritzker School of Molecular Engineering,
University of Chicago, Chicago, Illinois 60637, United States}
\affiliation{Materials Science Division and Center for Molecular Engineering, Argonne National Laboratory, Lemont, Illinois 60439, United States}
\affiliation{Department of Chemistry, University of Chicago, Chicago, Illinois 60637, United States}

\date{\today}

\begin{abstract}
Recently, we introduced the screened-exchange range-separated hybrid (SE-RSH) functional to account for spatially dependent dielectric screening in complex materials. The SE-RSH functional has shown good performance in predicting the electronic properties of a large variety of semiconductors and insulators, and of heterogeneous systems composed of building blocks with large dielectric mismatch. Here, we assess the performance of SE-RSH for oxide materials, including  antiferromagnetic transition-metal oxides. Through a comparison with other dielectric-dependent hybrid functionals, we demonstrate that SE-RSH yields improved predictions of dielectric constants and band gaps, bringing them into a closer agreement with experimental values. The functional also provides accurate values of magnetic moments of several oxides.
\end{abstract}

\maketitle


\section{\label{sec:level1}Introduction}
Metal oxide semiconductors and insulators are versatile materials used in numerous applications, ranging from energy conversion and storage to environmental technologies\cite{danish2020systematic, pazhamalai2024investigating, wang2017oxygen, naseem2021role, fine2010metal}. The broad applicability of metal oxide (MO) semiconductors  stems from their optoelectronic properties and the ability of metal cations to readily switch oxidation states under the presence of external fields, making  MOs particularly valuable for redox and charge transfer reactions. These characteristics have positioned these materials at the forefront of research in sustainable energy sources, including solar-driven water splitting\cite{kudo2009heterogeneous, osterloh2008inorganic}, photovoltaics\cite{jose2009metal}, and heterogeneous catalytic processes\cite{royer2011catalytic}.

From a theoretical standpoint, the description of MOs relevant to energy conversion technologies poses several challenges. For example, the study of heterogeneous systems at the atomistic level, such as the interface between an oxide semiconductor and an electrolyte present in photoelectrochemical cells, often  requires large supercells and advanced theoretical methods to describe light-matter interactions (e.g., many-body perturbation theory\cite{hedin1965new, martin2016interacting} or time-dependent techniques). However, such approaches can be computationally prohibitive for large systems. Consequently, approximate strategies are often employed, where density functional theory (DFT) is used to treat solid and liquid phases separately, followed by an \emph{a posteriori} description of their interaction\cite{tomasi2005quantum, lipparini2016perspective, andreussi2012revised}.

Even putting aside the complexities of materials heterogeneity, computing accurate properties of MOs using DFT remains challenging. 
Local and semi-local exchange and correlation energy functional (xc) functionals are known to severely underestimate the value of fundamental band gaps\cite{shishkin2007self}. Hybrid functionals, in which the exchange energy is obtained as a linear combination of exact Hartree-Fock and local exchange energies, have emerged as a more accurate alternative to (semi)-local functionls, with PBE0\cite{adamo1999toward} and HSE\cite{heyd2003hybrid,10.1063/1.2204597,krukau2006influence} being popular ones to
describe condensed systems.

Among hybrid functionals, dielectric-dependent (DD) hybrid functionals\cite{skone2016nonempirical, skone2014self, marques2011density, cui2018doubly, brawand2017performance}, such as DD-RSH-CAM\cite{chen2018nonempirical}, RSH\cite{skone2016nonempirical} and DDH\cite{skone2014self}, have been increasingly adopted to investigate the structural and electronic properties of pristine \cite{marques2011density, brawand2017performance} and defective\cite{gerosa2015defect} solids, liquids\cite{gaiduk2018electron,pham2017electronic} and also of several molecules\cite{brawand2017performance}. For solids, this class of functionals is defined  using the dielectric constant of the system. Despite the success of DD hybrid functionals in predicting the band gaps of many insulators and semiconductors, several studies\cite{liu2019assessing, das2019band}  have  pointed at remaining inaccuracies for bulk metal oxides. Specifically, the DD-RSH-CAM functional tends to overestimate the band gaps of specific oxides, especially those of challenging antiferromagnetic transition-metal monoxides, MnO, CoO and NiO\cite{liu2019assessing, chen2018nonempirical}. Alternative approaches include Koopmans-compliant functionals\cite{borghi2014koopmans, colonna2019koopmans, nguyen2018koopmans, colonna2022koopmans} and related functionals based on the Wannier–Koopmans Method (WKM)\cite{ma2016using, weng2017wannier, weng2018wannier}, both of which reliably predict band gaps for main-group covalent semiconductors. However, the performance of WKM for the band gaps of transition metal oxides with partially filled d-states is not fully satisfactory\cite{weng2020wannier}.

Recently, we proposed the SE-RSH functional\cite{zhan2023nonempirical}, designed to extend the applicability of dielectric-dependent hybrid functionals to complex heterogeneous systems. By explicitly incorporating a spatially dependent dielectric screening, SE-RSH can account for varying screening environments experienced by different electronic states, and can accurately describe both localized and delocalized states. Hence, SE-RSH is expected to capture the often mixed character of oxide materials' band edges,  to which both itinerant (arising from oxygen s and p orbitals) and localized (arising from metal cation d or f orbitals) states may contribute. 

In this work, we provide a comprehensive validation of the performance of the SE-RSH functional for various metal oxides, focusing on band gaps, dielectric constants, and magnetic moments. We obtain results in good agreement with experiments for systems ranging from simple oxide, such as alumina to challenging antiferromagnetic systems, such as  Co and Ni oxides.

The remainder of this paper is organized as followed. In Section \ref{sec:method}, we revisit the expression of SE-RSH and highlight key differences from previous dielectric-dependent hybrid functionals. Section \ref{sec:result} presents electronic properties computed using SE-RSH, and Section \ref{sec:conclusion} summarizes our findings.

\section{Method\label{sec:method}}
\subsection{General Hybrid Functionals for Complex Materials}
When defining range-separated hybrid functionals\cite{leininger1997combining, vydrov2006assessment, vydrov2006importance, yanai2004new, chai2008long, stein2009reliable, baer2010tuned}, the non-local exchange potential $v_{\mathrm{x}}(\mathbf{r, r'})$ that enters the Kohn-Sham (KS) Hamiltonian  under the generalized Kohn-Sham scheme takes the form:
\begin{equation}
    v_{\mathrm{x}}(\mathbf{r, r'}) = \alpha(\mathbf{r, r'})\Sigma_{\mathrm{x}}(\mathbf{r, r'}) + (1-\alpha_{lr})v_{\mathrm{x}}^{\mathrm{lr}}(\mathbf{r};\mu) + (1 - \alpha_{sr})v_{\mathrm{x}}^{\mathrm{sr}}(\mathbf{r};\mu),
\end{equation}
where $\Sigma_{\mathrm{x}}$ represents the Hartree-Fock (exact) exchange, while $v_{\mathrm{x}}^{\mathrm{lr}}$ and $v_{\mathrm{x}}^{\mathrm{sr}}$ denote the long-range (lr) and short-range (sr) components of the semilocal exchange, respectively. The mixing fraction $\alpha(\mathbf{r, r'})$ controlling the combination of exact and semilocal exchange is given by:
\begin{equation}
    \alpha(\mathbf{r, r'}) = \alpha_{lr} + (\alpha_{sr} - \alpha_{lr})\mathrm{erfc}(\mu|\mathbf{r-r'}|),
\end{equation}
where $\alpha_{lr}$ and $\alpha_{sr}$ determine the fraction of exact exchange in the lr and sr components, respectively, and $\mu$ is a screening parameter.

Recently, range-separated dielectric-dependent hybrid functionals (RS-DDH)\cite{skone2016nonempirical, chen2018nonempirical, cui2018doubly} have been proposed,  that yield accurate electronic properties of inorganic materials and molecular crystals by relating the parameters entering $\alpha(\mathbf{r, r'})$ to system-dependent dielectric properties. Specifically, in RS-DDH, $\alpha_{lr}$ is set to the inverse of the macroscopic dielectric constant $\varepsilon_\infty^{-1}$, while $\alpha_{sr}$  has been chosen equal to $0.25$\cite{skone2016nonempirical} or $1$(DD-RSH-CAM\cite{chen2018nonempirical}). However, this approach relies on global parameters and is not expected to perform accurately in systems composed of regions with significant dielectric mismatch.

\begin{figure}[htbp!]
\includegraphics[scale=0.27]{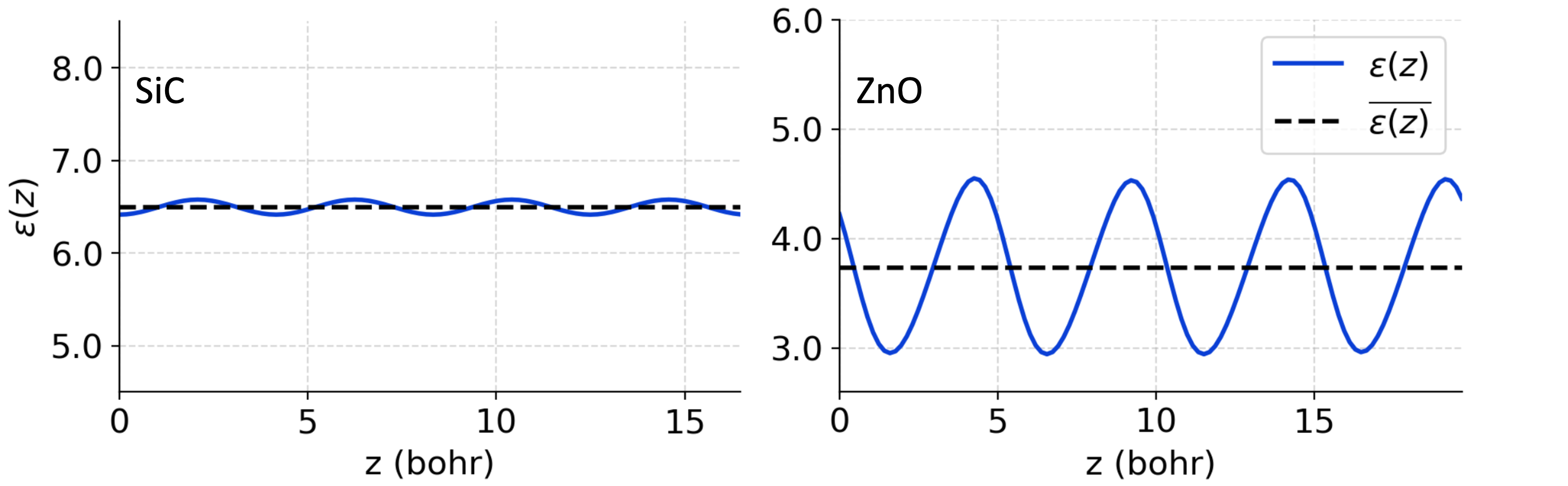}
\caption{\label{eps_comp}Comparison of the local dielectric function ($\varepsilon(z)$, average of the dielectric function $\varepsilon(\mathbf{r})$ in the $(x, y)$ plane) between SiC and ZnO.}
\end{figure}

\begin{figure}[htbp!]
\includegraphics[scale=0.28]{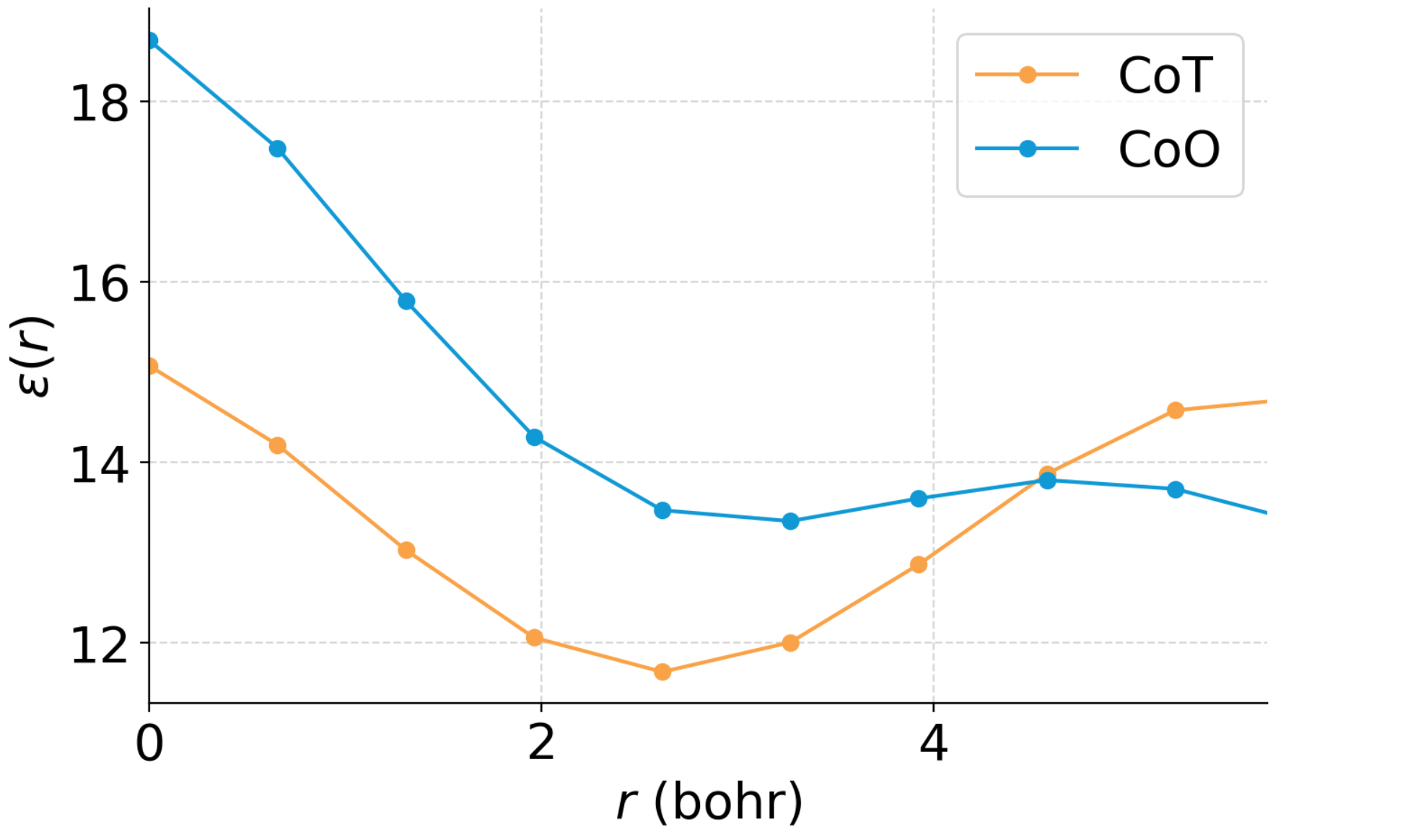}
\caption{\label{eps_co3o4}Comparison of the local dielectric function $\varepsilon(r)$, computed as the spherical average of $\varepsilon(\mathbf{r})$, centered on two nonequivalent Co ions in $\mathrm{Co_3O_4}$: $\mathrm{Co^{2+}}$ with tetrahedral oxygen coordination (CoT) and $\mathrm{Co^{3+}}$ with octahedral oxygen coordination (CoO).}
\end{figure}

To address the limitations of the DDH and RS-DDH functionals, we developed the screened-exchange range-separated hybrid (SE-RSH) functional, which provides a more general framework by incorporating local variations of the dielectric screening. The SE-RSH mixing fraction is defined as:
\begin{equation}
    \alpha^{\mathrm{SE-RSH}}(\mathbf{r, r'}) = \frac{1}{\sqrt{\varepsilon(\mathbf{r})\varepsilon(\mathbf{r'})}} + \left(1 - \frac{1}{\sqrt{\varepsilon(\mathbf{r})\varepsilon(\mathbf{r'})}}\right)\mathrm{erfc}\left(\mu(\mathbf{r})|\mathbf{r-r'}|\right). \label{sersh_equ}
\end{equation}
Here, $\varepsilon(\mathbf{r})$ and $\mu(\mathbf{r})$ represent a local dielectric function and local screening function, respectively; $\varepsilon(\mathbf{r})$ is determined from DFT calculations in finite electric field, by evaluating the shifts of Wannier function centers in the presence of a static field  to obtain the spatially dependent induced polarization.\cite{zheng2019dielectric, stengel2006accurate, souza2002first}. Fig.(\ref{eps_comp}) illustrates the distinct dielectric screening profiles of a covalently bonded semiconductor (SiC) and a metal oxide (ZnO). The significant ionic character of metal-oxygen bonds in ZnO leads to strong variations in the local charge density near each ion, resulting in a highly position-dependent polarizability. By explicitly incorporating $\varepsilon(\mathbf{r})$ rather than a single global $\varepsilon_{\infty}$ parameter, the SE-RSH functional effectively captures the local screening fluctuations and their impact on the electronic structure of the material.
This sensitivity to local screening is particularly valuable for systems with ions in distinct oxidation states. For instance, when using a DFT+U \cite{anisimov1991band} method to study $\mathrm{Co_3O_4}$, the presence of different oxidation states requires two separate Hubbard U parameters \cite{chen2011electronic}; however, as shown in Fig.(\ref{eps_co3o4}), the SE-RSH functional naturally accounts for the differences in screening around each ion.

\begin{figure}[htbp!]
\includegraphics[scale=0.2]{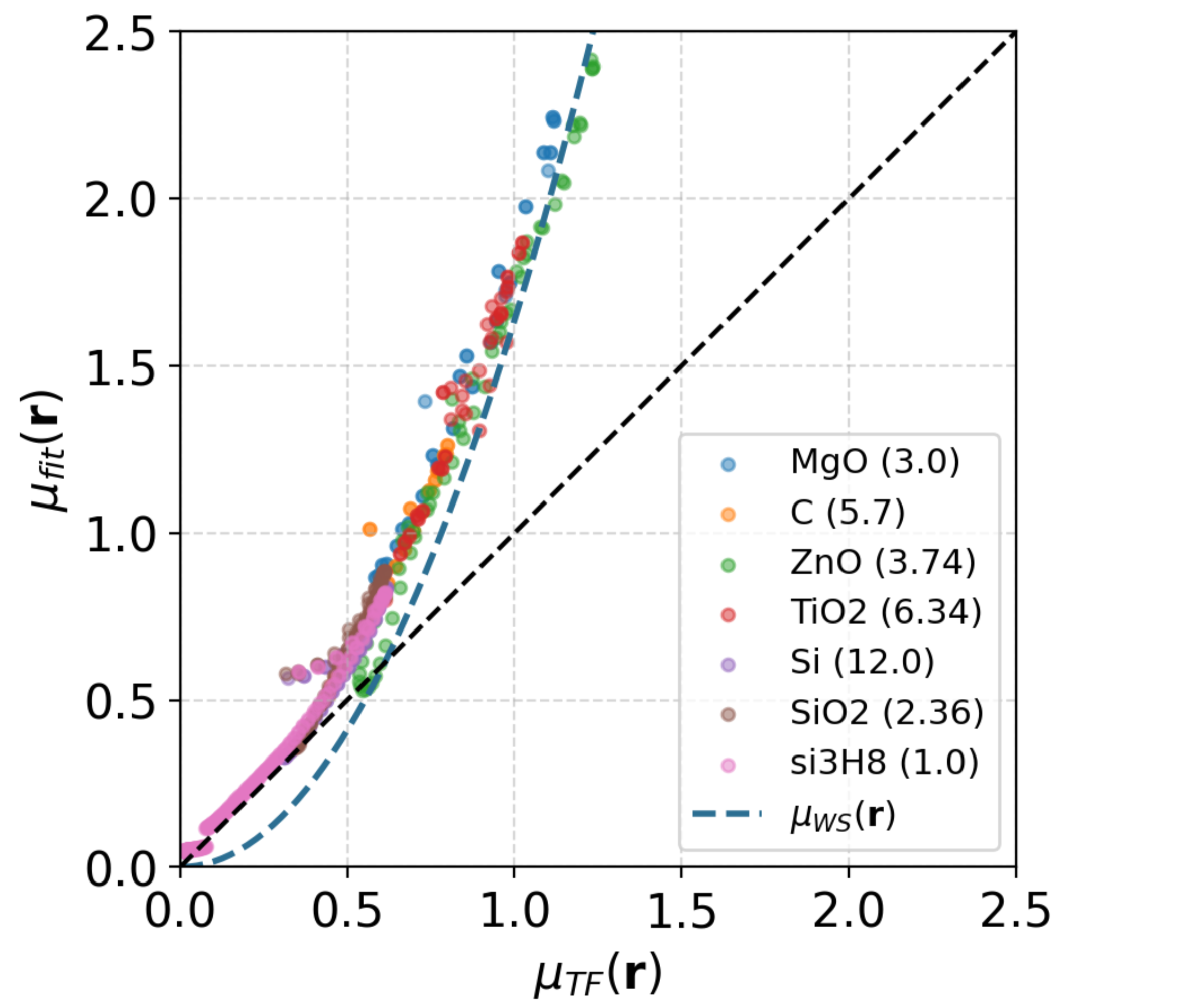}
\caption{\label{mu_fit}Comparison of $\mu_{\mathrm{fit}}(\mathbf{r})$ (see Eq.(\ref{equ_fit})) with $\mu_{\mathrm{WS}}(\mathbf{r})$ and $\mu_{\mathrm{TF}}(\mathbf{r})$ (see text) at randomly selected points in real space ,$\{\mathbf{r}\}$, for several materials with different band gaps and dielectric constants (values in parentheses).}
\end{figure}

The local screening function $\mu(\mathbf{r})$ generalizes the global $\mu$ parameter used in RS-DDH. Previous work by Skone et al.\cite{skone2016nonempirical} and Chen et al.\cite{chen2018nonempirical} determined $\mu$ by fitting a model dielectric function:
\begin{equation}
    \varepsilon^{-1}_{\mathrm{model}}(\mathbf{G}) = \varepsilon^{-1}_{\infty} + \left(1 - \varepsilon^{-1}_{\infty}\right)\left(1 - e^{-\frac{|\mathbf{G}|^2}{4\mu^2}}\right)
\end{equation}
to the long-range decay of the diagonal elements of the dielectric matrix $\varepsilon^{-1}(\mathbf{G, G'})$ computed using, for example, the linear response techniques proposed in Ref.\cite{wilson2008efficient} and implemented in the open-source $\texttt{WEST}$ (Without Empty STates) code\cite{govoni2015large, yu2022gpu}. This approach is formally equal to approximating the screened Coulomb interaction $W(\mathbf{r, r'})$ as:
\begin{eqnarray}
    W(\mathbf{r, r'}) &=& \int \varepsilon^{-1}(\mathbf{r, r''})v(\mathbf{r'', r'})d\mathbf{r''} \nonumber \\
    & \approx & \frac{\varepsilon^{-1}_{\infty}}{|\mathbf{r-r'}|} + \left(1 - \varepsilon^{-1}_{\infty}\right)\frac{\mathrm{erfc}(\mu|\mathbf{ r-r'}|)}{|\mathbf{r-r'}|}.
\end{eqnarray}
We extend this approximation by considering a generalized model $W(\mathbf{r, r'})$, assuming that a local screening function $\mu(\mathbf{r})$ is sufficient to capture the screening decay near a given point $\{\mathbf{r}\}$:
\begin{eqnarray}
    W(\mathbf{r, r'}) \approx \frac{\varepsilon^{-1}(\mathbf{r})}{|\mathbf{r-r'}|} + \left(1 - \varepsilon^{-1}(\mathbf{r})\right)\frac{\mathrm{erfc}(\mu(\mathbf{r})|\mathbf{ r-r'}|)}{|\mathbf{r-r'}|}.\label{equ_w}
\end{eqnarray}
This generalization leads to a spatially resolved model for the dielectric function, as the Fourier transform of $W(\mathbf{r, r'})$ in Eq.(\ref{equ_w}) with respect to $\mathbf{r'}$ is:
\begin{eqnarray}
    W(\mathbf{r, G'}) & = & \left[\varepsilon^{-1}(\mathbf{r}) + \left(1 - \varepsilon^{-1}(\mathbf{r})\right)\left(1 - e^{-\frac{|\mathbf{G'}|^2}{4\mu^2(\mathbf{r})}}\right)\right]e^{-i\mathbf{G'}\cdot\mathbf{r}} \frac{4\pi}{|\mathbf{G'}|^{2}} \\
     & = & \varepsilon^{-1}_{\mathrm{model}}(\mathbf{r, G'})v(\mathbf{G'}),
\end{eqnarray}
where $v(\mathbf{G'}) = 4\pi/|\mathbf{G'}|^{2}$. We can then determine $\mu(\mathbf{r})$ through a fitting procedure:
\begin{equation}
    \mu_{\mathrm{fit}}(\mathbf{r}) = \operatorname*{argmin}_{\mu(\mathbf{r})} \sum_{\mathbf{G'}}\left|\varepsilon^{-1}_{\mathrm{model}}(\mathbf{r, G'}) - \varepsilon^{-1}(\mathbf{r, G'})\right|^{2}.\label{equ_fit}
\end{equation}
However, the direct implementation of this protocol would be computationally demanding, requiring the calculations of the dielectric function at each $\{\mathbf{r}\}$ point in real space. To address this challenge, we approximate $\mu_{\mathrm{fit}}(\mathbf{r})$ using quantities that can be easily computed from first principles. Through an analysis of the value of $\mu_{\mathrm{fit}}(\mathbf{r})$ across diverse materials at randomly selected $\{\mathbf{r}\}$ points, we found that the following approximation is accurate:
\begin{equation}
\mu_{\mathrm{fit}}(\mathbf{r}) \approx \mathrm{max}\left(\mu_{\mathrm{WS}}(\mathbf{r}), \mu_{\mathrm{TF}}(\mathbf{r})\right)\label{mu_equ}
\end{equation}
as shown in Fig.(\ref{mu_fit}), where $\mu_{\mathrm{WS}}(\mathbf{r}) = \frac{1}{r_{s}(\mathbf{r})}=\left(\frac{4\pi n_v(\mathbf{r})}{3}\right)^{1/3}$,  $\mu_{\mathrm{TF}}(\mathbf{r}) = \frac{1}{2}k_{\mathrm{TF}} = \left(\frac{3n_v(\mathbf{r})}{\pi}\right)^{1/6}$, $n_v$ is the valence electron density, $r_s$ is the Wigner-Seitz radius, and $k_{\mathrm{TF}}$ is the Thomas–Fermi screening wavevector.
We note  that $\mu_{\mathrm{WS}}$ can be interpreted as providing a measure of how tightly electrons are packed in the material, while $\mu_{\mathrm{TF}}$ describes the collective response of the electron gas to external perturbations. By taking the maximum of the two values, $\mu_{\mathrm{fit}}$ ensures that the dominant screening mechanism in the system (whether local or collective) is adequately described.

\subsection{Computational Details}
The SE-RSH hybrid functional has been implemented in a modified version of the  $\texttt{Qbox}$\cite{gygi2008architecture} code, which employs a plane-wave basis set and norm-conserving pseudopotentials. We performed calculations for 13 metal oxides at the experimental geometry, using supercells containing between 128 and 256 atoms (see Supplementary Table \ref{structure} for details), and the $\Gamma$ point to sample the supercell Brillouin zone. We used optimized norm-conserving Vanderbilt pseudopotentials\cite{hamann2013optimized}; semicore states were included for alkaline earth metals and (post-)transition metals. The energy cutoff for the plane-wave basis set was set at 90 Ry. The local dielectric function  was determined  self-consistently using the SE-RSH functional: at the first iteration, we performed a DFT calculation at the PBE level ($\alpha(\mathbf{r, r'}) = 0$); then we set $\varepsilon(\mathbf{r})$ in Eq.(\ref{sersh_equ}) to be the local dielectric function from the previous iteration until $\varepsilon(\mathbf{r})$ and the total energy were converged.

The $\texttt{WEST}$\cite{govoni2015large, yu2022gpu} code was used to compute the dielectric matrix using $6\times N_{\mathrm{elec}}$ eigenpotentials for each system considered here, where $N_{\mathrm{elec}}$ is the number of electrons. The local screening function $\mu(\mathbf{r})$ used in SE-RSH was computed according to Eq.(\ref{mu_equ}).

\section{Electronic Structure of Oxides Using the SE-RSH Functional\label{sec:result}}
\subsection{Screening parameter and macroscpic dielectric constant\label{sec_param}}
\begin{figure}[h]
\includegraphics[scale=0.2]{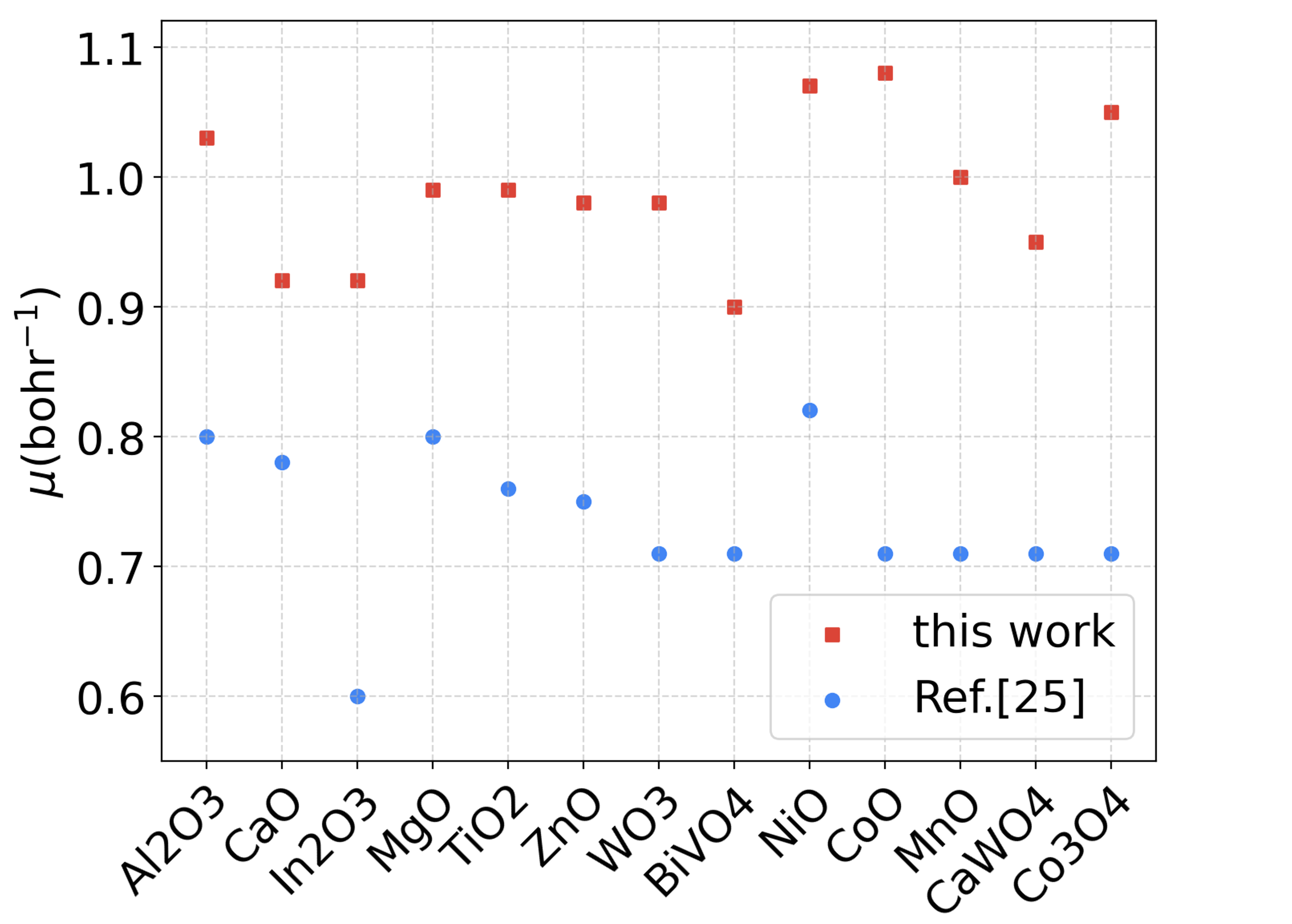}
\caption{\label{mu_comp}Comparison of global screening parameters $\mu$ from least-squares fits of the dielectric functions with those reported in Ref. \cite{chen2018nonempirical} (see Supplementary Table \ref{table_eps} for details).}
\end{figure}

\begin{figure}[h]
\includegraphics[scale=0.27]{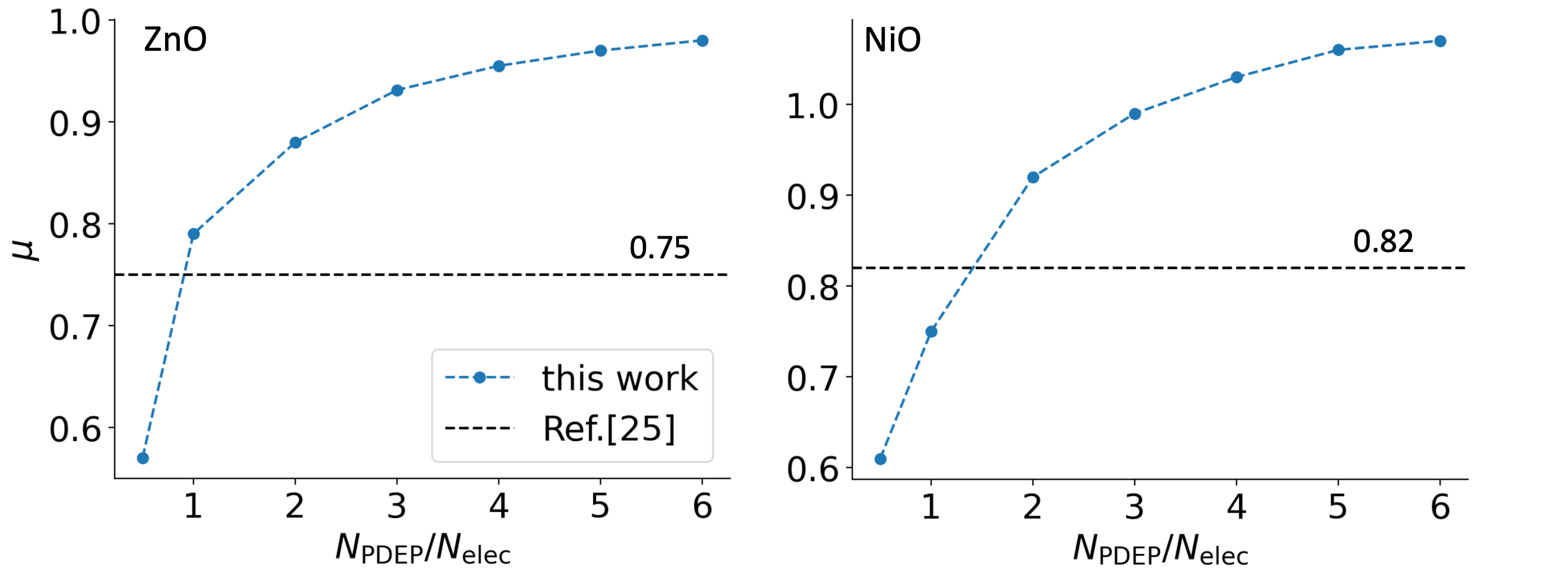}
\caption{\label{mu_pdep}The global screening parameters $\mu$ of ZnO and NiO, computed by fitting the dielectric functions, as a function of the ratio of the projective dielectric eigenpotential (PDEP)\cite{wilson2008efficient} basis set size to the number of electrons. For comparison, horizontal dashed lines indicate the $\mu$ values for both systems as reported in Ref.\cite{chen2018nonempirical} and used by DD-RSH-CAM.}
\end{figure}

We computed the global screening parameters $\mu$ for 13 oxides obtained by a least-squares fit to the dielectric functions and we compared our results with those of Ref.\cite{chen2018nonempirical} (see Fig.(\ref{mu_comp}) and Supplementary Table \ref{table_eps}).
Our calculated values differ from those of Ref.\cite{chen2018nonempirical}, most likely due to methodological differences in computing the dielectric matrix or the irreducible polarizability $\chi^{0}$. Chen et al. obtained the noninteracting Green's function $G$ through a perturbative expansion over the Kohn-Sham eigenstates, requiring a convergence of the results with respect to the number of unoccupied states; instead we employed the projective dielectric eigenpotential (PDEP) approach\cite{wilson2008efficient} implemented in the $\texttt{WEST}$ code\cite{govoni2015large, yu2022gpu}. The PDEP method computes the dielectric matrix without the explicit evaluation of empty electronic states, with the accuracy controlled solely by the size of the PDEP basis set ($N_{\mathrm{PDEP}}$). We verified the convergence of $\mu$ with respect to $N_{\mathrm{PDEP}}$, as demonstrated in Fig.(\ref{mu_pdep}). As we show below, the value of the screening parameter $\mu$ strongly influences the electronic structure predictions of the DD-RSH-CAM functional, making the methodological differences in computing the dielectric matrix particularly significant for accurate band gap predictions.

\begin{figure}[h]
\includegraphics[scale=0.30]{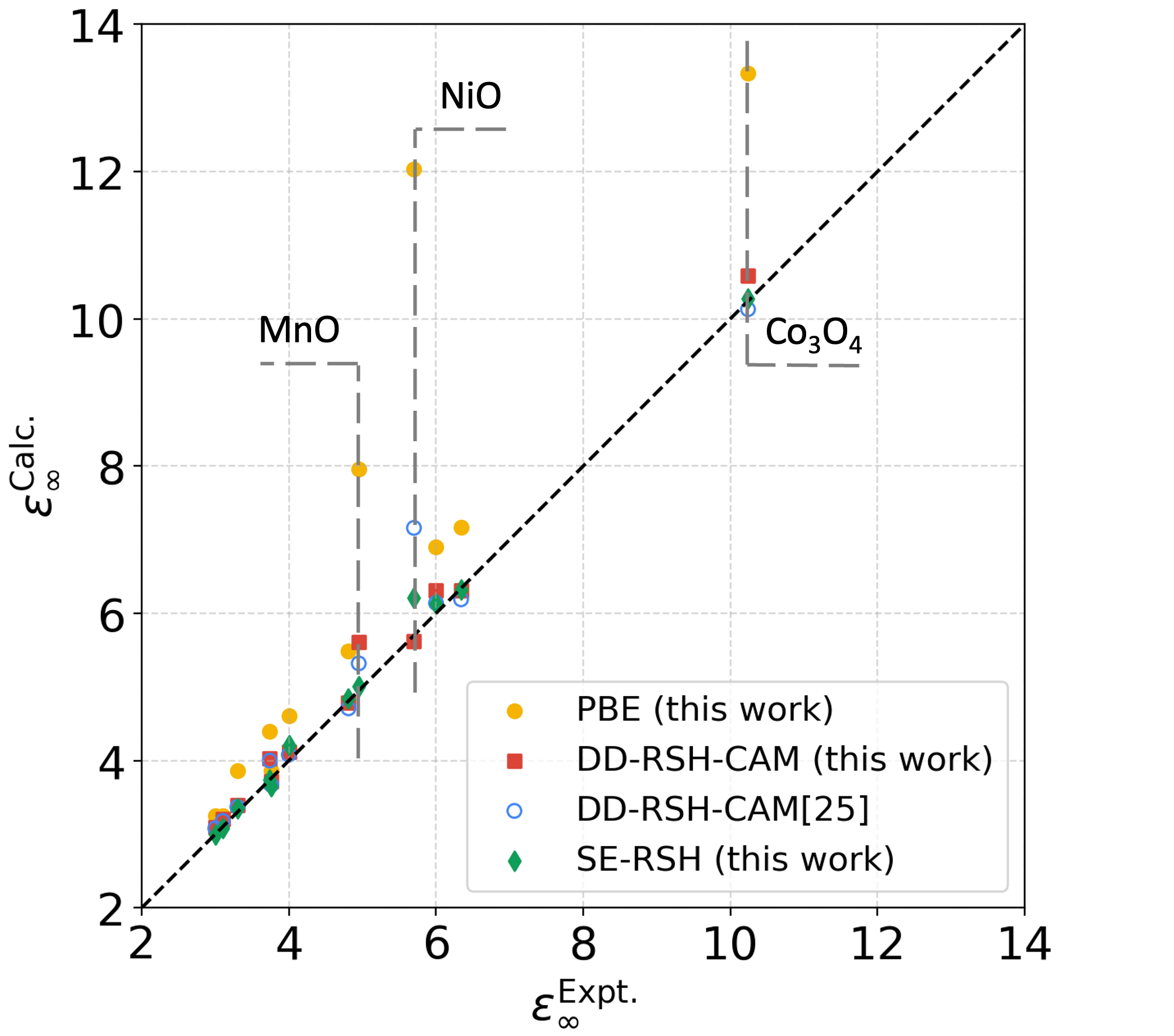}
\caption{\label{eps_comp_xc}Comparison of experimental macroscopic dielectric constants $\varepsilon_\infty^{\mathrm{Expt.}}$ with calculated values $\varepsilon_\infty^{\mathrm{Calc.}}$ using the finite-field method at the PBE level and self-consistently at the level of dielectric-dependent hybrid functionals (see Supplementary Table \ref{table_eps} for details). See text for the definition of the DD-RSH-CAM and SE-RSH functionals.}
\end{figure}

\begin{figure}[h]
\includegraphics[scale=0.35]{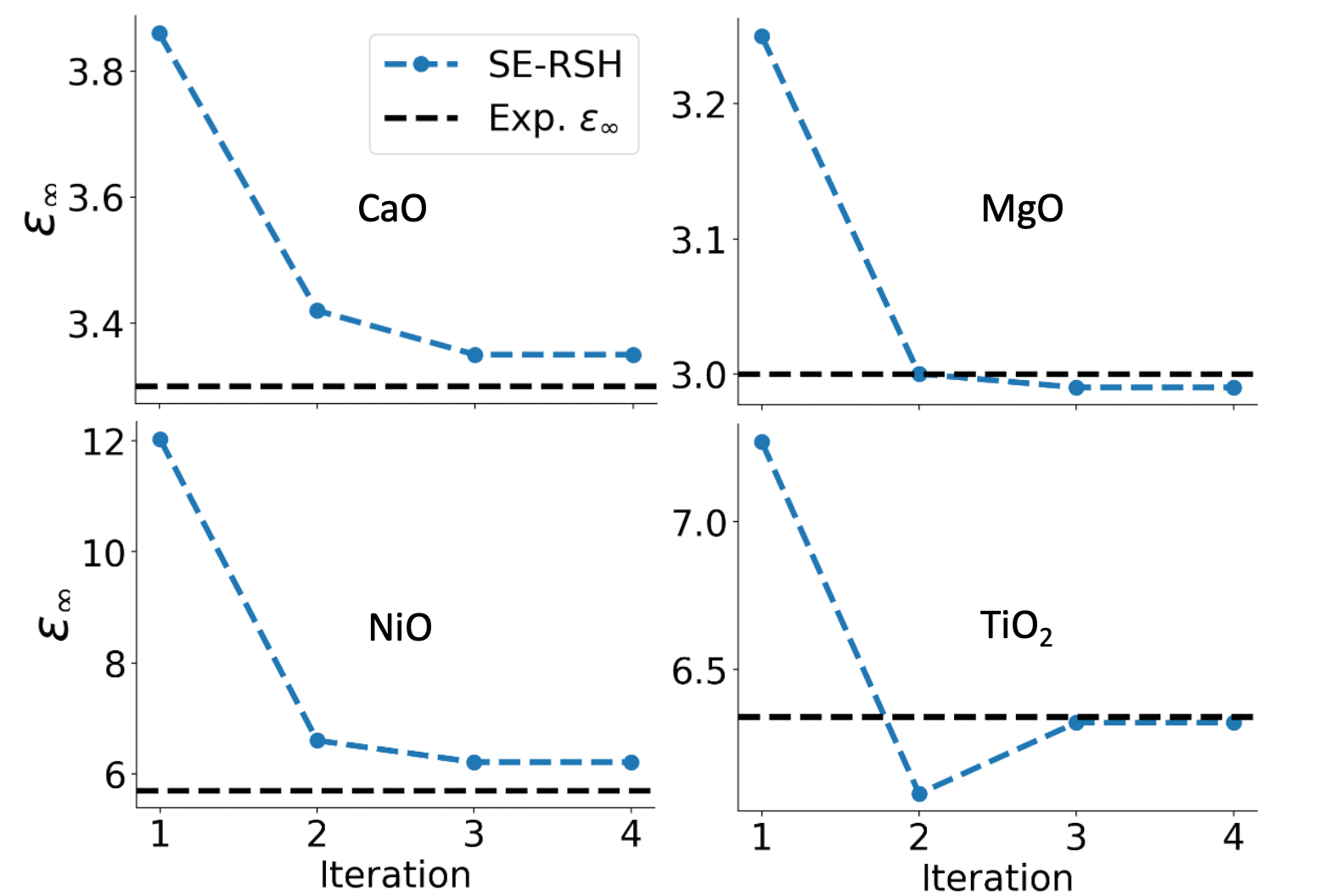}
\caption{\label{eps_iter}Macroscopic dielectric constants of CaO, MgO, NiO and $\mathrm{TiO_2}$, computed using the finite field method at the level of SE-RSH, as a function of the number of iteration of the self-consistent procedure. The horizontal dash lines denote experimental values.}
\end{figure}

Fig.(\ref{eps_comp_xc}) also presents macroscopic dielectric constants ($\varepsilon_{\infty}$) computed using both the finite field (FF) method at the PBE level and self-consistently at the hybrid functional level.
For DD-RSH-CAM calculations, we fixed $\mu$ to the values specified in Supplementary Table \ref{table_eps}. The finite field approach based on the PBE functional (FF@PBE)  systematically overestimates $\varepsilon_{\infty}$ compared to the experimental values, with particularly pronounced deviations for strongly correlated Mott insulators such as NiO and MnO. In contrast, FF@hybrid calculations significantly improve the accuracy of our results, with SE-RSH achieving a mean absolute percentage error (MAPE) of just 2.34\% (details are shown in Supplementary Table \ref{table_eps}). Additionally, we note that the self-consistent determination of dielectric constants using FF@hybrid typically converges within a few iterations, as shown in Fig.(\ref{eps_iter}).
\subsection{Band gap}
\begin{figure}[h]
\includegraphics[scale=0.30]{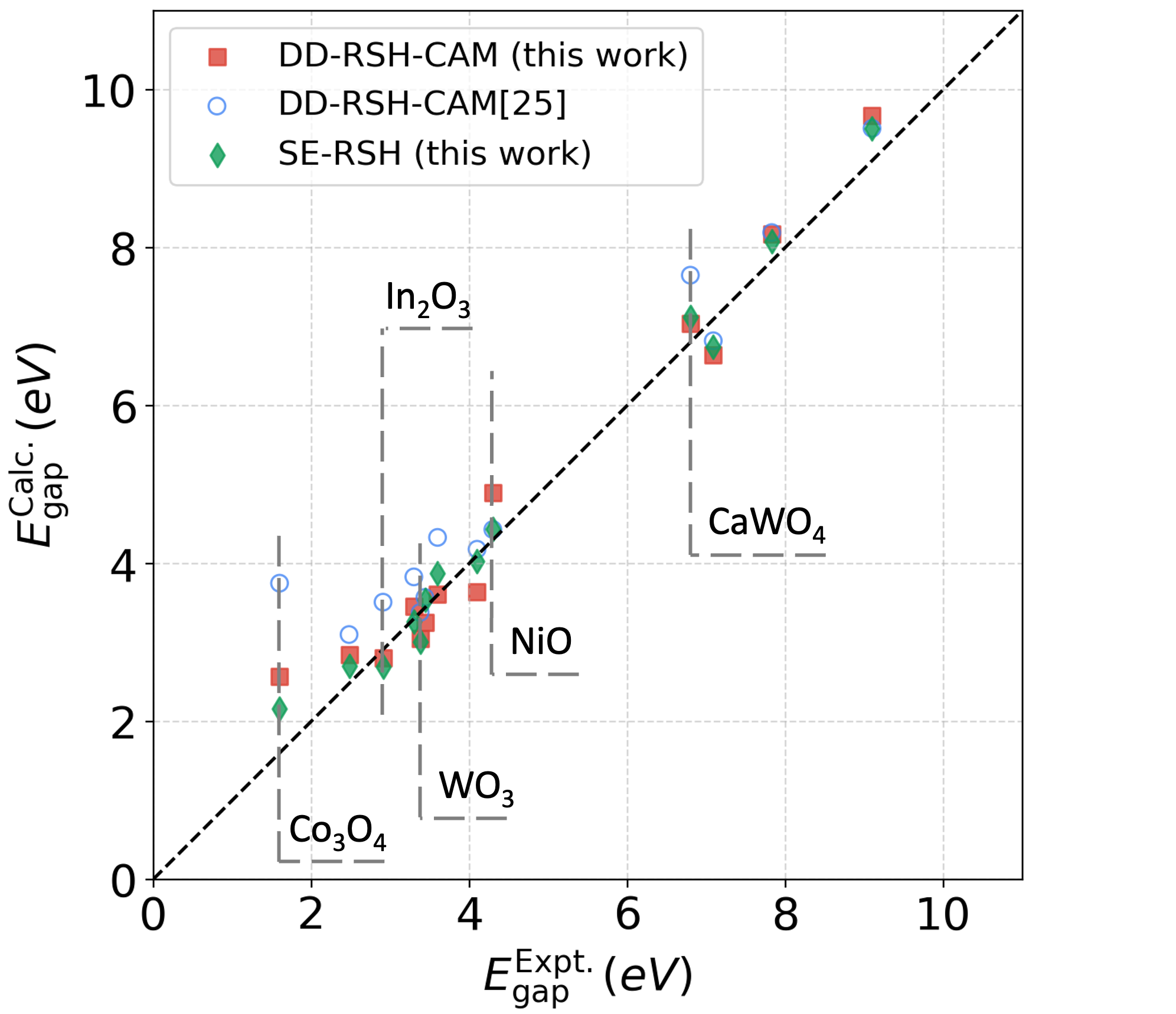}
\caption{\label{gap_comp}Fundamental band gaps $E_{\mathrm{gap}}^{\mathrm{Calc.}}$(eV) computed for 13 metal oxides, compared with experimental values $E_{\mathrm{gap}}^{\mathrm{Expt.}}$(eV) (see Supplementary Table \ref{table_gap} for details).}
\end{figure}

In Fig.(\ref{gap_comp}), we compare the band gaps obtained for oxides using the SE-RSH hybrid functional and other dielectric-dependent hybrid functionals, including the DD-RSH-CAM which utilizes $\frac{1}{\varepsilon_\infty}$ as the fraction of Fock exchange in the long-range part and a constant screening parameter $\mu$.
To evaluate the impact of the choice of $\mu$ on the performance of DD-RSH-CAM, we present results using both our own computed values (see computational details) and those from Ref.\cite{chen2018nonempirical}. Because our calculations do not include electron–phonon coupling, we subtract the value of the zero-phonon renormalization (ZPR) from computed band gaps, where available, for a comparison with experimental data. The ZPR values are also shown in Supplementary Table \ref{table_gap}.

The SE-RSH functional yields accurate band gaps for most of the systems studied, with a mean error (ME) of 0.09 eV and a mean absolute error (MAE) of 0.26 eV (see Supplementary Table \ref{table_gap} for details).  For some oxides, such as NiO and $\mathrm{Co_3O_4}$, whose band gaps are typically overestimated by DD-RSH-CAM, the SE-RSH tends to yield predictions closer to experiments. This improved accuracy arises primarily from the SE-RSH’s ability to capture variations in the local dielectric screening of metal oxides, even though both functionals are formally equivalent when $\varepsilon(\mathbf{r})$ and $\mu(\mathbf{r})$ are constant. We note that for certain oxides, such as $\mathrm{Al_2O_3}$, $\mathrm{MgO}$, $\mathrm{CaWO_4}$ and $\mathrm{Co_3O_4}$, SE-RSH slightly overestimates the band gap; however, no ZPR values for these oxides are available and hence they were not subtracted. Additionally, for $\mathrm{WO_3}$ and $\mathrm{Co_3O_4}$, the discrepancy between SE-RSH and experimental values exceeds 10\%. This large difference may be partly attributed to experimental data collected from thin films\cite{shinde2006supercapacitive} or under conditions with high surface sensitivity\cite{meyer2010charge}, which might lead to differences from bulk values.

It is worth noting that some of the DD-RSH-CAM results reported here differ from those of Ref.\cite{chen2018nonempirical}. Specifically, our calculated DD-RSH-CAM band gaps are smaller for a few compounds (including $\mathrm{In_2O_3}$, $\mathrm{TiO_2}$, ZnO), a difference due to our computed values for $\mu$ (see Section \ref{sec_param}). The only exception is NiO, for which our DD-RSH-CAM calculations predict a larger gap, primarily due to differences in the calculated dielectric constant. To better understand the effect of $\mu$, we performed additional calculations using experimental dielectric constants. For NiO, we find a significantly overestimated band gap of 5.7 eV when using $\mu = 0.82$ from Ref.\cite{chen2018nonempirical}, suggesting this $\mu$ value may be inappropriate. While our computed $\mu$ values overall improve the accuracy of DD-RSH-CAM, this functional still yields  overestimated gaps  for NiO and $\mathrm{Co_3O_4}$, indicating that an accurate treatment of the spatially varying dielectric screening is crucial for these systems.

\subsection{Magnetic moment of antiferromagnetic TMO}
Finally, we assess the performance of SE-RSH in predicting magnetic moments of antiferromagnetic transition metal oxides. Table \ref{table_mm} presents calculated magnetic moments using different functionals alongside experimental values. The PBE functional underestimates magnetic moments, consistent with its known tendency to over delocalize electrons. Hybrid functionals, including SE-RSH, produce larger magnetic moments in closer agreement with experimental data. Notably, the predicted magnetic moments show relatively low sensitivity to the specific hybrid functional employed, suggesting that the specific treatment of exact exchange plays a less critical role in determining magnetic properties than it does for band gaps and dielectric constants, at least for the materials studied here.

\begin{table}[h]
\caption{\label{table_mm} Magnetic moments (in $\mathrm{\mu B}$/atom) calculated by several functionals as well as from available experimental values.}
\begin{ruledtabular}
\begin{footnotesize}%
\begin{tabular}{cdddc}
    & \multicolumn{1}{c}{\text{PBE}}
    & \multicolumn{1}{c}{\text{DD-RSH-CAM}}
    & \multicolumn{1}{c}{\text{SE-RSH}}
    & \multicolumn{1}{c}{\text{Expt.}}
    \\
\hline
NiO & 1.19 & 1.67 & 1.66 & 1.64\cite{aplerin1962magnetic}, 1.90\cite{roth1958magnetic}\\
MnO & 4.08 & 4.46 & 4.52 & 4.58\cite{cheetham1983magnetic}\\
CoO & 2.24 & 2.76 & 2.71 & 2.46\cite{csiszar2005controlling}\\
\end{tabular}
\end{footnotesize}%
\end{ruledtabular}
\end{table}

\section{Conclusions\label{sec:conclusion}}
In summary, we have presented a comprehensive assessment of the SE-RSH functional for metal oxides by computing band gaps, dielectric constants and magnetic moments, and comparing our results with those of  other range-separated dielectric-dependent hybrid functionals and with experiments. The key innovation of the SE-RSH functional lies in its spatially resolved treatment of the dielectric screening, incorporating both local dielectric and screening functions in determining the exact-exchange mixing fraction. This approach enables an accurate treatment of the local variations in the dielectric screening characteristic of metal oxides. We find that  SE-RSH can  predict fundamental band gaps, dielectric constants, and magnetic moments in close agreement with experiments. Our results underscore the importance of self-consistently computing the macroscopic dielectric constant at the hybrid functional level, as this corrects the significant overestimation observed at the PBE level of theory. Furthermore, our analysis reveals that different  implementations can yield substantially different screening parameters $\mu$, helping to explain previously reported band gap overestimates by the DD-RSH-CAM functional for certain systems. Work is in progress to extending the application of the SE-RSH functional to investigate oxide-oxide and oxide-water interfaces\cite{Matthew2025}, where the accurate treatment of spatially varying dielectric screening is particularly crucial for understanding electrochemical and photocatalytic processes.

\begin{acknowledgments}
The authors thank Yu Jin for valuable discussions. This work was supported by DOE/BES through the Computational Materials Science Center Midwest Integrated Center for Computational Materials (MICCoM). The computational resources were provided by the University of Chicago’s Research Computing Center and the National Energy Research Scientific Computing Center (NERSC).
\end{acknowledgments}

\appendix

\bibliography{main}

\makeatletter\@input{r.tex}\makeatother
\end{document}



\title{Supporting Information: Dielectric-Dependent Range-Separated Hybrid Functional Calculations for Metal Oxides}

\author{Jiawei Zhan}
\affiliation{Pritzker School of Molecular Engineering,
University of Chicago, Chicago, Illinois 60637, United States}
\author{Marco Govoni}
\affiliation{Pritzker School of Molecular Engineering,
University of Chicago, Chicago, Illinois 60637, United States}
\affiliation{Materials Science Division and Center for Molecular Engineering, Argonne National Laboratory, Lemont, Illinois 60439, United States}
\affiliation{Department of Physics, Computer Science, and Mathematics, University of Modena and Reggio Emilia, Modena, 41125, Italy}
\author{Giulia Galli}
\email{gagalli@uchicago.edu}
\affiliation{Pritzker School of Molecular Engineering,
University of Chicago, Chicago, Illinois 60637, United States}
\affiliation{Materials Science Division and Center for Molecular Engineering, Argonne National Laboratory, Lemont, Illinois 60439, United States}
\affiliation{Department of Chemistry, University of Chicago, Chicago, Illinois 60637, United States}

\date{\today}

\maketitle

\clearpage
\renewcommand{\theequation}{S\arabic{equation}}
\renewcommand{\thetable}{S\arabic{table}}
\renewcommand{\thefigure}{S\arabic{figure}}
\setcounter{equation}{0}
\setcounter{table}{0}
\setcounter{figure}{0}

\begin{table}[h]
\caption{The space group, lattice parameters, and the number of atoms in the supercell used for the calculations of 13 oxides.}\label{structure}
\begin{ruledtabular}
\begin{footnotesize}%
\begin{tabular}{cccccccccc}
& Space group & a($\mathring{A}$) & b($\mathring{A}$) & c($\mathring{A}$) & $\alpha(^{\circ})$ & $\beta(^{\circ})$ & $\gamma(^{\circ})$ & Reference & \# of atoms \\
\hline
$\mathrm{Al_2O_3}$ & $R\overline{3}c$& 4.760 & 4.760 & 12.990 & 90 & 90 & 90 & \cite{kondo2008structural} & 240\footnotemark[1]  \\
$\mathrm{CaO}$ &    $Fm\overline{3}m$ & 4.803 & 4.803 & 4.803 & 90 & 90 & 90 & \cite{smith1968low} & 256\footnotemark[1] \\
$\mathrm{In_2O_3}$ & $Ia\overline{3}$ & 10.117  & 10.117  & 10.117  & 90 & 90 & 90 & \cite{marezio1966refinement} & 160\footnotemark[1]\\
$\mathrm{MgO}$ &   $Fm\overline{3}m$ & 4.207  & 4.207  & 4.207  & 90 & 90 & 90 & \cite{smith1968low} & 256\footnotemark[1]   \\
$\mathrm{TiO_2}$ & $P4_2/mnm$ & 4.593 & 4.593 & 2.958 & 90 & 90 & 90 & \cite{burdett1987structural} & 192\footnotemark[1]  \\
$\mathrm{ZnO}$ &  $P6_3mc$ & 3.250  & 3.250  & 5.203 & 90 & 90 & 90 & \cite{karzel1996lattice} & 128\footnotemark[1]    \\
$\mathrm{WO_3}$ &  $P2_1/c$ & 7.301 & 7.539 & 7.690 & 90 & 90.89 & 90 & \cite{woodward1995structure} & 256\footnotemark[1]  \\
$\mathrm{BiVO_4}$ & $I2/b$ & 5.190 & 5.090 & 11.700 & 90 & 90 & 90.38 & \cite{wang2020role} & 192\footnotemark[1]  \\
$\mathrm{NiO}$ &  $Fm\overline{3}m$ & 4.171 &  4.171 & 4.171 & 90 & 90 & 90 & \cite{bartel1971exchange} & 256\footnotemark[2]\\
$\mathrm{CoO}$ &  $Fm\overline{3}m$ & 4.254 &  4.254 & 4.254 & 90 & 90 & 90 & \cite{carey1991preparation} & 256\footnotemark[2]\\
$\mathrm{MnO}$ &  $Fm\overline{3}m$ & 4.445 &  4.445 & 4.445 & 90 & 90 & 90 & \cite{barrett1964solid} & 256\footnotemark[2] \\
$\mathrm{CaWO_4}$ & $I4_1/a$ & 5.249 & 5.249 & 11.393 & 90 & 90 & 90 & \cite{de2021electronic} & 144\footnotemark[1]  \\
$\mathrm{Co_3O_4}$ & $Fd\overline{3}m$ &  8.084 & 8.084 & 8.084 & 90 & 90 & 90 & \cite{smith1973structure} & 224\footnotemark[1] \\
\end{tabular}
\footnotetext[1]{Supercell sizes were chosen to ensure convergence of the computed fundamental band gap values (see Fig. \ref{pic_convergence}).}
\footnotetext[2]{We determined the appropriate size of the supercell by comparing the computed fundamental band gaps with those from primitive cell calculations using an 8×8×8 k-point mesh with a global dielectric-dependent hybrid functional. We chose $\alpha = 1/\varepsilon_\infty$, where $\varepsilon_\infty$ is the experimental macroscopic dielectric constant, for this comparison. Results with 256-atom supercells yield band gaps of 3.47 eV, 3.20 eV, and 3.16 eV for NiO, CoO, and MnO, respectively, compared to 3.51 eV, 3.17 eV, and 3.09 eV obtained from the primitive cell calculations performed using \texttt{Quantum Espresso}.}
\end{footnotesize}%
\end{ruledtabular}
\end{table}

\begin{figure}[h]
\includegraphics[scale=0.45]{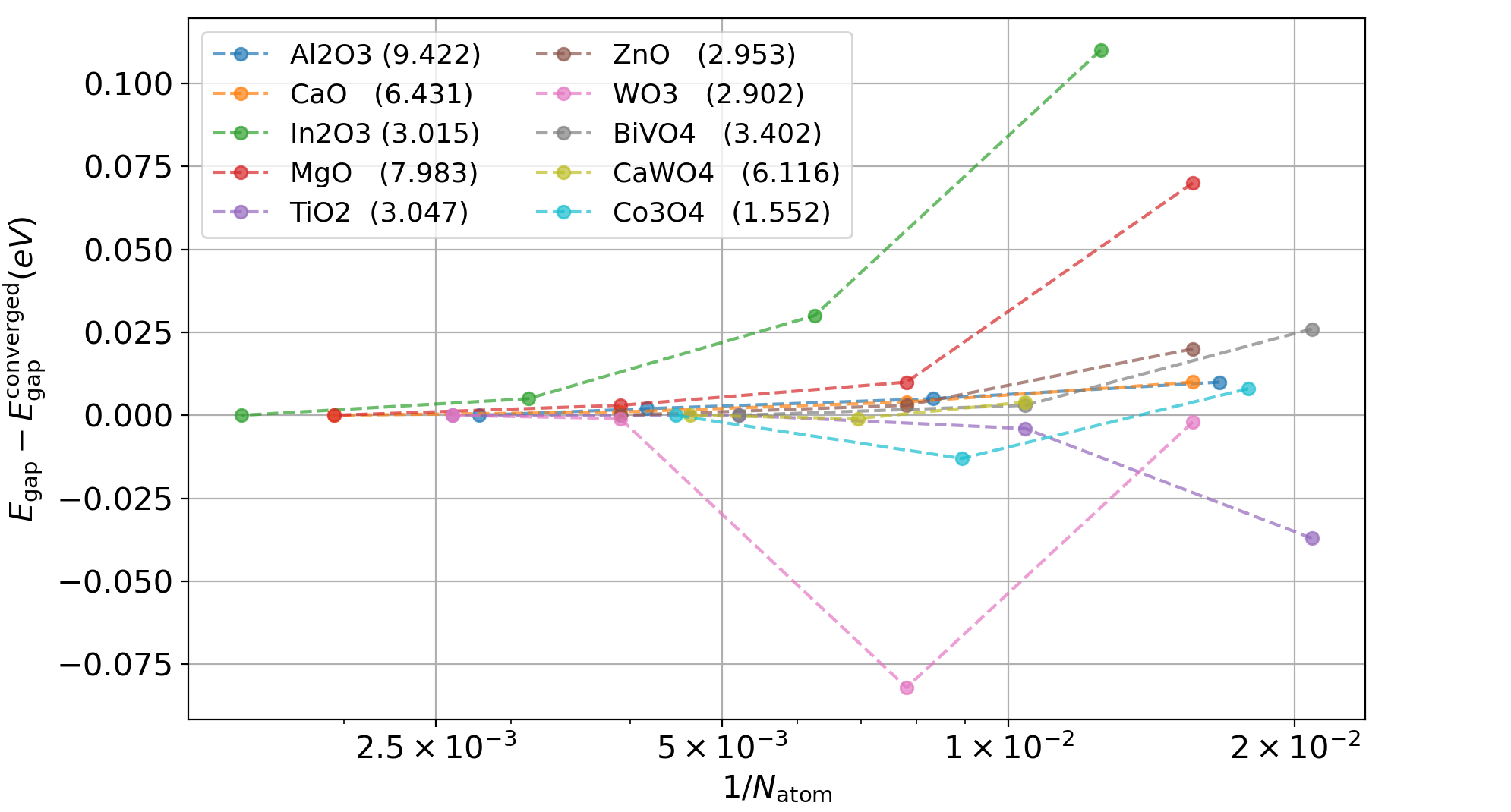}
\caption{\label{pic_convergence} Convergence of band gaps with increasing number of atoms ($N_{\mathrm{atom}}$) in the supercell, computed using the global dielectric-dependent hybrid functional ($\alpha=1/\varepsilon_\infty$, where $\varepsilon_\infty$ is the experimental macroscopic dielectric constant). Values in parentheses in the inset indicate the converged $E_{\mathrm{gap}}$ in eV.}
\end{figure}

\begin{table}[h]
\caption{The global screening parameters $\mu$ obtained in this work via least-squares fit to the dielectric functions compared with the values reported in Ref.\cite{chen2018nonempirical} (columns 2 and 3). The macroscopic dielectric constants $\varepsilon_\infty$ calculated using the finite-field method at the PBE level and self-consistently at the level of dielectric-dependent hybrid functionals are given in columns 4-7 and the experimental results in column 8. The mean error (ME), mean absolute error (MAE), and mean absolute percentage error (MAPE) relative to experimental dielectric constants are also reported.}\label{table_eps} 
\begin{ruledtabular}
\begin{footnotesize}%
\begin{tabular}{cddcdddc}
& \multicolumn{2}{c}{$\mu(\mathrm{bohr}^{-1})$} 
    & \multicolumn{5}{c}{$\varepsilon_\infty$} \\
    \cline{2-3}\cline{4-8}
    & \multicolumn{1}{c}{\text{DD-RSH-CAM}}
    & \multicolumn{1}{c}{\text{DD-RSH-CAM\cite{chen2018nonempirical}}}
    & \multicolumn{1}{c}{\text{PBE}}
    & \multicolumn{1}{c}{\text{DD-RSH-CAM}}
    & \multicolumn{1}{c}{\text{DD-RSH-CAM\cite{chen2018nonempirical}}}
    & \multicolumn{1}{c}{\text{SE-RSH}}
    & \multicolumn{1}{c}{\text{Expt.}}
    \\
\hline
$\mathrm{Al_2O_3}$ & 1.03  & 0.80  & 3.25  & 3.21  & 3.18  & 3.08  & 3.10  \\
$\mathrm{CaO}$ &     0.92  & 0.78  & 3.86  & 3.39  & 3.37  & 3.35  & 3.30  \\
$\mathrm{In_2O_3}$ & 0.92  & 0.60  & 4.61  & 4.12  & 4.08  & 4.21  & 4.00  \\
$\mathrm{MgO}$ &     0.99  & 0.80  & 3.25  & 3.09  & 3.08  & 2.99  & 3.00  \\
$\mathrm{TiO_2}$ &   0.99  & 0.76  & 7.17  & 6.31  & 6.19  & 6.32  & 6.34  \\
$\mathrm{ZnO}$ &     0.98  & 0.75  & 4.40  & 4.03  & 4.00  & 3.74  & 3.74  \\
$\mathrm{WO_3}$ &    0.98  & 0.71\footnotemark[1]  & 5.49  & 4.79  & 4.71\footnotemark[2] & 4.84  & 4.81  \\
$\mathrm{BiVO_4}$ &  0.90  & 0.71\footnotemark[1]  & 6.90  & 6.31  & 6.14\footnotemark[2] & 6.13  & 6.00  \\
$\mathrm{NiO}$ &     1.07  & 0.82  & 12.03 & 5.62  & 7.16  & 6.21  & 5.70  \\
$\mathrm{CoO}$ &     1.08  & 0.71\footnotemark[1]  & -- & 5.63  & 5.29\footnotemark[2] & 4.97  & 5.30  \\
$\mathrm{MnO}$ &     1.00  & 0.71\footnotemark[1]  & 7.96  & 5.61  & 5.32\footnotemark[2] & 5.01  & 4.95  \\
$\mathrm{CaWO_4}$ &  0.95  & 0.71\footnotemark[1]  & 3.86  & 3.72  & 3.67\footnotemark[2] & 3.65  & 3.76  \\
$\mathrm{Co_3O_4}$ & 1.05  & 0.71\footnotemark[1]  & 13.34 & 10.59 & 10.13\footnotemark[2] & 10.27 & 10.24 \\
$\mathrm{ME}$ & & & 1.43 & 0.17 & 0.16 & 0.04 \\
$\mathrm{MAE}$ & & & 1.43 & 0.19 & 0.23 & 0.12 \\
$\mathrm{MAPE(\%)}$ & & & 25.84 & 3.96 & 4.60 & 2.34 \\
\end{tabular}
\end{footnotesize}%
\end{ruledtabular}
\footnotetext[1]{$\mu$ is set to 0.71, in accordance with the recommendation in Ref.\cite{chen2018nonempirical}.}
\footnotetext[2]{$\varepsilon_\infty$ is determined self-consistently using the finite-field method @ DD-RSH-CAM with $\mu=0.71$}.
\end{table}

\begin{table}[h]
\caption{\label{table_gap} Fundamental band gaps $E_g$(eV) computed for 13 metal oxides; we subtracted from  each theoretical value the zero-phonon renormalization (ZPR) $E_{\mathrm{ZPR}}$ (column~7) when available. Most experimental values are obtained from the optical gap extrapolated to 0 K, with the exciton binding energy added, when available. The ME, MAE, and MAPE values relative to $E_{\mathrm{Expt.}}$ are also reported.}
\begin{threeparttable}
\begin{ruledtabular}
\begin{scriptsize}%
\begin{tabular}{cdddddc}
    & \multicolumn{1}{c}{\text{PBE}}
    & \multicolumn{1}{c}{\text{DD-RSH-CAM}}
    & \multicolumn{1}{c}{\text{DD-RSH-CAM\cite{chen2018nonempirical}}}
    & \multicolumn{1}{c}{\text{SE-RSH}}
    & \multicolumn{1}{c}{\text{Expt.}}
    & \multicolumn{1}{c}{$E_{\mathrm{ZPR}}$}
    \\
\hline
$\mathrm{Al_2O_3}$ & 6.25 & 9.67 & 9.51 & 9.51 & 9.10\textsuperscript{b} & \\
$\mathrm{CaO}$     & 3.26 & 6.64 & 6.82 & 6.74 & 7.09\textsuperscript{c} & 0.35\cite{engel2022zero} \\
$\mathrm{In_2O_3}$ & 1.21 & 2.80 & 3.51 & 2.69 & 2.91\textsuperscript{d} \\
$\mathrm{MgO}$     & 4.80 & 8.17 & 8.19 & 8.08 & 7.83\textsuperscript{e} \\
$\mathrm{TiO_2}$   & 1.61 & 3.46 & 3.83 & 3.26 & 3.30\textsuperscript{f} & 0.35\cite{engel2022zero} \\
$\mathrm{ZnO}$     & 0.69 & 3.25 & 3.57 & 3.54 & 3.44\textsuperscript{g} & 0.17\cite{engel2022zero} \\
$\mathrm{WO_3}$    & 1.51 & 3.05 & 3.38\textsuperscript{a} & 3.01 & 3.38\textsuperscript{h} & 0.4\cite{ping2013optical} \\
$\mathrm{BiVO_4}$  & 1.83 (1.16\textsuperscript{p}) & 3.52 (2.85\textsuperscript{p}) & 3.77\textsuperscript{a} (3.10\textsuperscript{p}) & 3.37 (2.70\textsuperscript{p}) & 2.48\textsuperscript{i} & 0.38\cite{wiktor2017comprehensive} \\
$\mathrm{NiO}$     & 0.72 & 4.90 & 4.43 & 4.44 & 4.30\textsuperscript{j} & 0.25\cite{abdallah2024quasiparticle} \\
$\mathrm{CoO}$     & \text{metal} & 3.61 & 4.33\textsuperscript{a} & 3.88 & 3.60\textsuperscript{k}& 0.28\cite{abdallah2024quasiparticle}  \\
$\mathrm{MnO}$     & 0.59 & 3.64 & 4.18\textsuperscript{a} & 4.03 & 4.1\textsuperscript{l} & 0.3\cite{abdallah2024quasiparticle} \\
$\mathrm{CaWO_4}$  & 4.13 & 7.04 & 7.65\textsuperscript{a} & 7.12 & 6.8\textsuperscript{m} \\
$\mathrm{Co_3O_4}$ & 0.31 & 2.57 & 3.75\textsuperscript{a} & 2.16 & 1.6\textsuperscript{n} \\
$\mathrm{ME}$      &  -2.51 &  0.13 &  0.49 &  0.09\\
$\mathrm{MAE}$     &  2.51  &  0.37 &  0.53 &  0.26\\
$\mathrm{MAPE(\%)}$ & 59.2 & 11.2  & 19.3  &  7.4\\
\end{tabular}
\end{scriptsize}%
\end{ruledtabular}
\begin{tablenotes}
    \item{\textsuperscript{a}Band gaps computed with DD-RSH-CAM using $\varepsilon_\infty$ and $\mu$ from Table \ref{table_eps};
    \textsuperscript{b}Ref.\cite{french1990electronic}, optical band gap extrapolated to 0 K;
    \textsuperscript{c}Ref.\cite{whited1973exciton} at 85 K;
    \textsuperscript{d}Optical band gap extrapolated to 0 K\cite{irmscher2014nature} plus exciton binding energy\cite{feneberg2016many};
    \textsuperscript{e}Ref.\cite{whited1973exciton} at 85 K;
    \textsuperscript{f}Ref.\cite{tezuka1994photoemission};
    \textsuperscript{g}Ref.\cite{mang1995band} at 6 K;
    \textsuperscript{h}Ref.\cite{meyer2010charge};
    \textsuperscript{i}Ref.\cite{cooper2015indirect};
    \textsuperscript{j}Ref.\cite{sawatzky1984magnitude} at 473 K;
    \textsuperscript{k}Ref.\cite{gvishi1972hall} via conductivity measurement, referring to 0K;
    \textsuperscript{l}Ref.\cite{kurmaev2008oxygen};
    \textsuperscript{m}Exciton reflectance peak at 295 K \cite{grasser1975optical} plus exciton binding energy\cite{zhang1998electronic};
    \textsuperscript{n}Ref.\cite{shinde2006supercapacitive}, optical band gap at 300 K.
    \textsuperscript{p}Spin-orbit coupling and nuclear quantum effects are also considered\cite{wiktor2017comprehensive}.
    }
\end{tablenotes}
\end{threeparttable}
\end{table}

\clearpage

\bibliography{main}